\documentclass[final, runningheads]{llncs}
\usepackage[T1]{fontenc}
\usepackage{graphicx}
\newcommand{\itadata}{\footnotesize \textsl{ITADATA2024: The 3$^{\text{rd}}$ Italian Conference on Big Data and Data Science}}
\usepackage{amsmath}
\usepackage{fancyhdr}
\usepackage[labelfont=bf]{caption}
\usepackage{wrapfig}
\usepackage{array}
\usepackage{tikz}
\usepackage{xcolor}
\usepackage{amsfonts}
\usepackage{enumerate}
\usepackage{pifont}
\usepackage{eqparbox}
\usepackage{subcaption}
\usepackage{setspace}
\usepackage{eqparbox}
\usepackage{subcaption}
\usepackage{multirow}
\usepackage{makecell}
\usepackage{float}

\usepackage{amsthm}
\usepackage{ulem}
\usetikzlibrary{3d}
\fancypagestyle{empty}{%
  \fancyhf{}
  \fancyhead[C]{\itadata}
  \fancyhead[R]{\includegraphics[width=.05\textwidth]{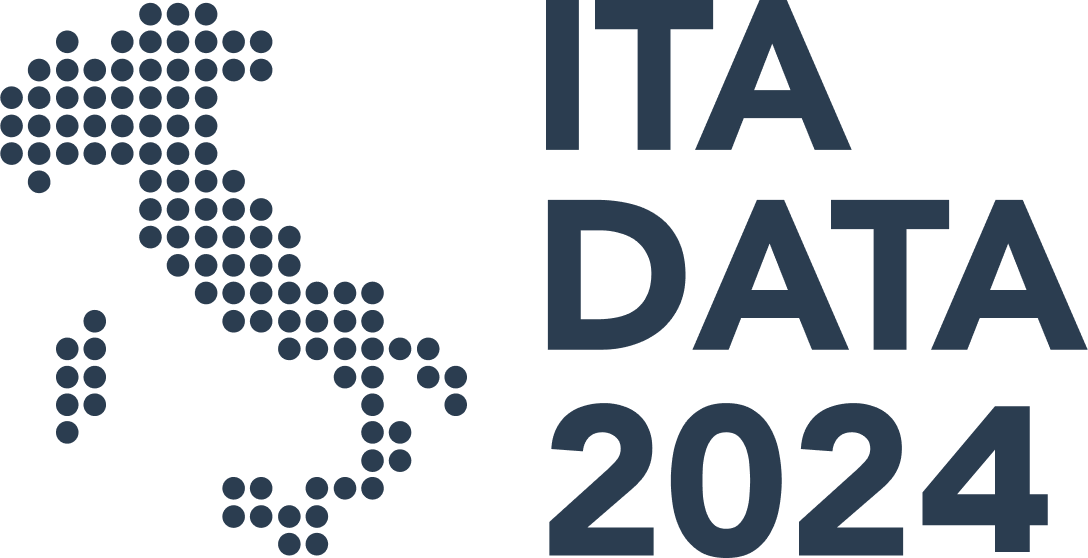}}
}

\begin{document}
\title{Space of Data through the Lens of \\ Multilevel Graph
}
%
%
\author{Marco Caputo 
\and
Michele Russo 
\and
Emanuela Merelli 
}

\authorrunning{F. Author et al.}
%
\institute{CINI Consortium - Data Science Lab \\ University of Camerino, Camerino 62032, IT \\
\email{name.surname@unicam.it}\\
}
\maketitle              
\begin{abstract}

This work seeks to tackle the inherent complexity of \textit{dataspaces} by introducing a novel data structure that can represent datasets across multiple levels of abstraction, ranging from local to global. We propose the concept of a multilevel graph, which is equipped with two fundamental operations: contraction and expansion of its topology. This multilevel graph is specifically designed to fulfil the requirements for incremental abstraction and flexibility, as outlined in existing definitions of \textit{dataspaces}. Furthermore, we provide a comprehensive suite of methods for manipulating this graph structure, establishing a robust framework for data analysis. While its effectiveness has been empirically validated for unstructured data, its application to structured data is also inherently viable. Preliminary results are presented through a real-world scenario based on a collection of dream reports.
\keywords{Graph Contraction \and Data Science \and Graph Analytics}
\end{abstract}

\section{Introduction}

%

In the contemporary era of big data and increasingly complex information systems, there is a growing need for sophisticated methods to represent and analyze large-scale datasets derived from diverse real-world domains. These datasets often encapsulate hidden relationships that, when uncovered, can yield significant insights, thereby contributing to human well-being and environmental~\mbox{sustainability}. Typically, raw data are converted into structured or semi-structured formats, such as data sources, making them suitable for integration within \mbox{\textit{dataspaces}}.

Although numerous definitions exist within the literature, a \textit{dataspace} is commonly defined as a collection of data sources (participants) and the interrelationships that connect them~\cite{Mrityunjay11,Hutterer23}.
It has also been defined as a method that allows for the coexistence of different data sources and formats, serving as a flexible
framework for organizing these relationships without forcing integration~\cite{Franklin05}.
Stream systems and sensors, as data sources, producing highly unstructured data formats contribute to \textit{dataspaces} in a manner that renders the unification of data representation a particularly desirable goal.

This study aims to address the complexity of \textit{dataspaces} by proposing a novel data structure, the multilevel graph, which represents datasets at multiple levels of abstraction, ranging from the local (raw data) to the global (\textit{dataspace}), with data sources functioning as an intermediary level.
The mapping process used to construct a data source from raw data typically involves data cleaning, normalisation, and feature selection techniques, ultimately resulting in a labelled graph that represents the structure of the input data.
We introduce the multilevel graph as a data structure equipped with two operations: contraction and expansion of its topology. The concept of a data space emerges as a robust framework for organising and representing large datasets. It offers a unified perspective on heterogeneous data sources while facilitating the integration of diverse data formats~\cite{Atzori23}.
Furthermore, the concept of graph data space emerges when the original graph reaches maximum contraction, where nodes (graphs) are replaced by points, and arcs are represented as proximity relations between these points~\cite{Merelli13,Rasetti16,Johnson13}.

Preliminary results highlighting the strengths and weaknesses of the proposed approach are presented using a real-world scenario derived from a collection of dream reports. The contraction operation effectively removes redundancies and noise by encapsulating global information in a more abstract and concise form, while preserving the ability to trace back to the original details. However, the computational complexity involved in graph manipulation remains a significant challenge, warranting further investigation.

\subsection{Related Work}
    Deriving a graph-like structure from unstructured data sources and proceeding to analysis supported
    by the means of graph analytics, is already a well-explored methodology in existing literature. Text written in
    natural language and, more specifically, dream narratives, present a unique challenge for
    data representation and analysis due to their subjective nature, varied content, and potential for hidden
    psychological insights, as has been analysed by Mota et al. with a statistical approach~\cite{Mota14}.
    Although the analysis is complete and interesting for the results, the applied methodology required a priori knowledge
    regarding the characteristics of the clusters used to classify behaviours. In our approach, the analysis starts without
    neither the personal information of the involved people nor the features regarding the classes of behaviour.
    We use the results of Mota et al. to interpret our semantic context.

    While these studies relate to our work in terms of data representation techniques, other works in the field of data
    integration made use of a multilevel paradigm to represent graph-like data spaces at different levels of
    abstraction, with particular concern for traceability~\cite{Mota16}. However, the transformation process used to
    produce such frameworks aims to produce a semantic reinterpretation of the data, providing a final
    abstract view that can only be obtained through semantic equivalences among levels, either manually assigned or
    automatically inferred through external resources.
    Our structure differs in terms of the information extraction focus, as we aim to provide a more general
    tool for the analysis of the topological aspects of the data space itself.
    Moreover, the multilevel graph maintains characteristics of determinism and independence from external sources of data.

    The ability of the multi-level graph to flexibly transition between different levels of granularity mirrors the
    properties of data spaces required by proposed agenda~\cite{Franklin05}, with potential for incremental query integration
    in a pay-as-you-go fashion, offered by the possibility of contraction and decontraction of nodes.
    While abstract data sources could provide limited base functionalities, like key-word search, decontractions could
    provide more sophisticated operations, such as relational-style queries.

\section{Multilevel Graph}

In this section, we introduce the formal definitions and basic properties of the multilevel graph.
Intuitively, a multilevel graph is a hierarchy of related graphs, where each level represents a different abstraction of
the same data.
The hierarchy is constructed upon the fundamental raw information in the base-level graph by iteratively
contracting nodes of each level into \textit{supernodes}, following the well-known concept of contraction
in graph theory.
Along with contraction, the operation of decontracting a {\it supernode} into a graph is one of the key features of the
multilevel graph which provides traceability. For such reason, we first introduce the concept of a graph equipped with
a decontracting operation, which we call a \textit{Decontractible Graph}.

\begin{definition}[Decontractible Graph]
        A \textbf{decontractible graph} is a quadruple $G = (V, E, dec_V, dec_E)$, where:
        \begin{itemize}
            \item $V$ is a set of elements called \textbf{supernodes};
            \item $E \subseteq V \times V$ is a set of ordered couples of supernodes, called \textbf{superedges};
            \item $dec_V : V \rightarrow \mathcal{G}_D$ is a function such that $dec_V(v) = (\mathcal{V}_v,
                \mathcal{E}_v, dec_{\mathcal{V}_v}, dec_{\mathcal{E}_v})$ is a decontractible graph represented by the
                supernode $v$;
            \item $dec_E : E \rightarrow \mathcal{P}(\mathcal{V} \times \mathcal{V})$ with $\mathcal{V} = \bigcup_{v \in V}\mathcal{V}_v$,
                is a function such that $\forall e = (u, v)$ with $u \neq v$, $dec_E(e) \subseteq$ $\{(a, b)$ $\mid$ $a \in \mathcal{V}_u$ $\wedge$
                $b \in \mathcal{V}_v\}$ is a set of edges represented by the superedge $e$.
        \end{itemize}
\end{definition}

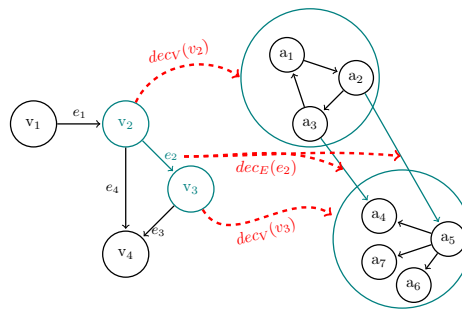
\begin{wrapfigure}{r}{0.5\textwidth}
 \vspace{-16pt}
      \centering
      \vspace{-20pt}
      \resizebox{!}{4cm}{
\begin{tikzpicture}
  [mynode/.style={draw, thick, circle, minimum size=10mm, inner sep=0mm},
    ->,shorten >=1pt,auto,node distance=2cm, thick,main node/.style={circle,draw, minimum size=10mm, inner sep=0mm},
    little node/.style={circle,draw, minimum size=7.5mm, inner sep=0mm},
    every edge/.style={draw, thick, ->}]

  \node[main node] (A) {v$_1$};
  \node[main node, teal] (B) [right of=A] {v$_2$};
  \node[main node, teal] (C) [below right of=B] {v$_3$};
  \node[main node] (D) [below left of=C] {v$_4$};

  \path[every node/.style={font=\sffamily\small}]
    (A) edge node [above] {$e_1$} (B)
    (B) edge[teal] node [right, name=e2] {$e_2$} (C)
    (C) edge node [below] {$e_3$} (D)
    (B) edge node [left] {$e_4$} (D);

  \begin{scope}[shift={(6,1)}]
    \draw[teal] (0,0) circle (1.5cm);
    \node[little node] (X) at (-0.5,0.5) {a$_1$};
    \node[little node] (Y) at (1,0) {a$_2$};
    \node[little node] (Z) at (0,-1) {a$_3$};
    \path[every node/.style={font=\sffamily\small}]
      (X) edge (Y)
      (Y) edge (Z)
      (Z) edge (X);
  \end{scope}

  \begin{scope}[shift={(8,-2.5)}]
    \draw[teal] (0,0) circle (1.5cm);
    \node[little node] (T) at (-0.5,0.5) {a$_4$};
    \node[little node] (U) at (1,0) {a$_5$};
    \node[little node] (V) at (0.25,-1) {a$_6$};
    \node[little node] (W) at (-0.5,-0.5) {a$_7$};
    \path[every node/.style={font=\sffamily\small}]
      (U) edge (T)
      (U) edge (W)
      (U) edge (V);
  \end{scope}

  \path[teal]
    (Z) edge[teal] (T)
    (Y) edge[teal] (U);

  \draw[dashed, line width=1.5pt, red] (B) to[out=65, in=145] node[midway, above, sloped, red] {$dec_{V}(v_2)$} (4.5,1);
  \draw[dashed, line width=1.5pt, red] (e2) to[out=0, in=155] node[midway, below, sloped, red] {$dec_{E}(e_2)$} (6.75,-1);
  \draw[dashed, line width=1.5pt, red] (e2) to[out=0, in=175] (8,-0.7);
  \draw[dashed, line width=1.5pt, red] (C) to[out=300, in=145] node[midway, below, sloped, red] {$dec_{V}(v_3)$} (6.5,-2);
\end{tikzpicture}
}
      \vspace{-17pt}
      \caption{ \small A local decontraction in a decontractible graph.}
      \label{fig:dec-graph-example}
         \vspace{-30pt}
\end{wrapfigure}

In the definition, as well as from now on, we will use the calligraphic notation (i.e. $\mathcal{V}$, $\mathcal{E}$,
\ldots) to indicate sets of {\it supernodes} and {\it superedges} obtainable through decontractions and, thus, located
at a lower level of the decontractible graph considered.
Specifically, the notation $G_v = (\mathcal{V}_v$, $\mathcal{E}_v)$ indicates the decontacted graph $dec_V(v)$ and
$\mathcal{E}_e$ the set of edges $dec_E(e)$.

It is worth noting that it is possible to use an attribute-based notation\\ 
as an alternative to the functional one to describe the characteristic properties of {\it supernodes} and {\it superedges} that make a graph decontractible. 
In this way, similar to traditional graphs, a decontractible graph can be represented as a couple $G = (V, E)$ where nodes
and edges are endowed with a special attribute $dec$ such that $v.dec = dec_V(v)$ for each $v \in V$ and
$e.dec = dec_E(e)$ for each $e \in E$.

Considering the recursive nature of the definition of a decontractible graph, if a {\it supernode} $v$ can be represented by
a decontractible graph $G_v$, then the nodes of $G_v$ will, in turn, be {\it supernodes}. Moreover, since decontractible graphs are effective graphs, all definitions given on standard graphs continue to be equivalently used for
decontractible graphs.

\noindent Compared to decontraction, the contraction relation intuitively allows decontractible graphs to correlate in the opposite sense.
While decontraction enables the construction of graphs, a definition of contraction seeks to establish the conditions that
must be valid on the couples of decontractibile graphs to make one an abstract representation of the other.

\begin{definition}[Contraction of a Decontractible Graph] \\
        Let $G = (V, E, dec_V, dec_E)$ be a decontractible graph, the decontractible graph
        $G\mathcal{'} = (\mathfrak{V}, \mathfrak{E}, dec_{\mathfrak{V}}, dec_{\mathfrak{E}})$ is a \textbf{contraction}
        of $G$ if and only if the following conditions are satisfied:
            \begin{enumerate}[(i)]
                \item the set $\{V_\alpha \mid \alpha \in \mathfrak{V} , \, dec_\mathfrak{V}(\alpha) = (V_\alpha, E_\alpha)\}$ is a partition of $V$;
                \item the set $(\{E_\alpha \mid \alpha \in \mathfrak{V} , \, dec_{\mathfrak{V}}(\alpha) = (V_\alpha, E_\alpha)\} \setminus \{ \emptyset \}) \cup
                    \{ dec_{\mathfrak{E}}(\epsilon) \mid \epsilon \in \mathfrak{E}\}$ is a partition of $E$.
            \end{enumerate}
    \end{definition}

Informally, the definition imposes that decontractions of {\it supernodes} and {\it superedges} in $G'$ must not be empty and that
the decontractions of the {\it supernodes} must be a partition of the {\it supernodes} of $G$. The condition of a partition
of the {\it superedges} naturally follows from the definition of decontractible graph.
In this definition, as well as henceforth, we will use the fraktur notation (i.e. $\mathfrak{V}$, $\mathfrak{E}$,
\ldots) to indicate sets of {\it supernodes} and {\it superedges} obtainable through
contractions and, thus, located at a higher level of the decontractible graph considered.
It is important to note that, given a contraction $G'$ of the decontractible graph $G$, all the information
necessary to compute the structure of the nodes and edges of $G$ can be derived from $G'$.
Specifically, let $G' = (\mathfrak{V}, \mathfrak{E})$ be a contraction of $G$, then from the definition we have

\begin{equation*}
        G = \left(\bigcup_{\alpha \in \mathfrak{V}} V_\alpha ,
            \left( \bigcup_{\alpha \in \mathfrak{V}} E_\alpha \cup \bigcup_{\epsilon \in \mathfrak{E}}{dec_{\mathfrak{E}}(\epsilon)} \right),
            dec_V, dec_E \right)
\end{equation*}

where $dec_V$ is obtained by union of the functions $dec_{V_\alpha}$ for each $\alpha \in \mathfrak{V}$,
and $dec_E$ is obtained by union of the functions $dec_{E_\epsilon}$ for each $\epsilon \in \mathfrak{E}$ combined with
all the mappings between {\it superedges} in $\bigcup_{\epsilon \in \mathfrak{E}}{dec_{\mathfrak{E}}(\epsilon)}$ and their
decontractions.
Therefore, let $\mathcal{G}_D$ be set of decontractible graphs, we can define the unary operator
$.^D : \mathcal{G}_D \rightarrow \mathcal{G}_D$ as the \textit{complete decontraction}
operator, which, given a decontractible graph $G$ returns the decontractible graph $G^D$ obtained by decontracting
all the {\it supernodes} and {\it superedges} of the input graph.

\noindent The next concept of {\it contraction function} helps us to define the multilevel graph.

\begin{definition}[Contraction Function] \newline
    A \textbf{contraction function} $f_C : \mathcal{G}_D \rightarrow \mathcal{G}_D$ is a function that, given a decontractible
    graph $G$, produces a new decontractible graph $f_C(G) = G'$ which is a contraction of $G$.
\end{definition}

A contraction function thus represents a specific contraction scheme where the domain and codomain coincide and are
represented by the set of decontractible graphs.
It produces contractions of the input decontractible graphs according to specific logic defined by the function itself.
A contraction function is closed within the set of decontractible graphs. This means that
it is possible to compose multiple contraction functions in sequence starting from a given decontractible graph.
Throughout this article, the terms \textit{contraction function} and \textit{contraction scheme} will be used
interchangeably.

\begin{definition}[Natural Transformation of a Graph] \newline
We define a \textbf{natural transformation function} $\eta$ as a function that, given a standard graph $H$
produces the corresponding decontractible graph \\ $G = (V, E, dec_V, dec_E)$ with the following properties:
\begin{enumerate}[(i)]
    \item \eqmakebox[things][l]{$dec_V(v) = (\emptyset, \emptyset)$}
    $ \begin{aligned}[t]
      \forall v\in V
      \end{aligned} $
    \item \eqmakebox[things][l]{$dec_E(e) = \emptyset$}
    $ \begin{aligned}[t]
      \forall e\in E
      \end{aligned} $
    \item $H$ and $G$ are isomorphic
\end{enumerate}
\end{definition}

Applying the natural transformation function on a graph means transforming a given standard graph into an
isomorphic decontractible graph with no possible decontractions.
It can be observed that these properties ensure that the obtained decontractible graph cannot be a contraction of any
other decontractible graph.

From the proposed definitions, the concept of the multilevel graph can finally be formally defined indicating an initial
graph and a stack of contraction schemes, making the concept of a hierarchy of related graphs naturally follow.

\, \vspace{-10pt}
\begin{wrapfigure}{r}{0.5\textwidth}
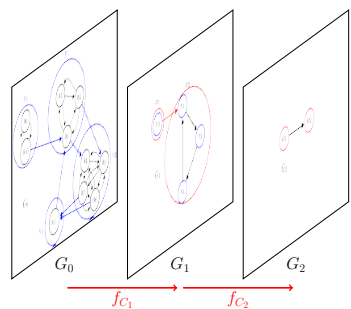

      \centering
      \vspace{-6pt}
      \include{tikzpics/mlg_example}
      \vspace{-0pt}
      \caption{ \small Representation of a multilevel graph of height 2.}
      \label{fig:mlg_example}
      \vspace{-40pt}
\end{wrapfigure}

\begin{definition}[Multilevel Graph] \newline
A \textbf{multilevel graph} $M$ is a pair $(G, \Gamma)$ \\ where:
\begin{itemize}
    \item $G = (V, E)$ is a graph;
    \item $\Gamma$ is a sequence $\langle f_{C_1}, f_{C_2}, \ldots, f_{C_k} \rangle$ of contraction functions.
\end{itemize}
\end{definition}

We use the notation $G_i$ to denote the contractible graph placed at the $i$-th level of the hierarchy.
Thus, the graph $G_0 = \eta(G)$ can be obtained from the natural transformation function $\eta$ applied to the standard
graph $G$, while the remaining graphs $G_i$ can be defined indicating that $f_{C_i}(G_{i-1}) = G_i$ for each
$i \in \{1, \ldots, k\}$, or equivalently, $G_k = (f_{C_k} \circ f_{C_{k-1}} \circ \ldots \circ f_{C_1} \circ \eta)(G)$.
For these reasons, given a multilevel graph $M = (G,\Gamma)$ with $\Gamma = \langle f_{C_1}, f_{C_2}, \ldots, f_{C_k}
\rangle$, the function that computes its contractible graph at the $k$-th level, $G_k$, can be described
through the following recursive definition:

\begin{equation*}
    con(M, k) =
    \left\{
    \begin{aligned}
        &f_{C_k}(con(M, k-1)) && \text{if } k > 0\\
        &\eta(G)  && \text{if } k = 0
    \end{aligned}
    \right\}
    = G_k
\end{equation*}

Let $\mathcal{M}$ be the set of multilevel graphs, we define the height function \newline
$h : \mathcal{M} \rightarrow \mathbb{N}$, such that $h(M) = k$, where $M = (G, \Gamma)$ and $k$ is the number of
contraction functions in $\Gamma$.

\section{Multilevel Graph Data Analysis}

The data analysis approach proposed in this work, called {\it Multilevel Graph Data Analysis (MGDA)}, is based on the
application of the multilevel graph data structure to a graph-like space of data, possibly derived from unstructured data sources.

This process of mapping raw data into a more concise and abstract relational representation can be achieved by
performing a preliminary phase of data cleaning, which establishes the core relevant information to be represented.

Normalisation is then applied, which allows the representation of the data in a standardised relational format (i.e. a graph).
This part of the process establishes the base structure of the space of data, defining the nature of the relationships between
the data points and, thus, the topological characteristics of the structure.
In the context of the multilevel analysis, the base structure of the space is represented by the initial graph $G$
of the multilevel graph $M = (G, \Gamma)$.

Feature detection and selection techniques are subsequently applied to the space of data, which allows the identification of the underlying
fundamental spatial information to be represented in upper levels.
The point of this phase is to minimize the overall complexity of the space, by removing redundant and noisy
information about the relationships between the elements and by highlighting the topological features of interest.
Feature detection can be applied recursively, this means that the result of feature detection at a certain level can
be used to detect new features at the next level.
Note that resulting representations at upper levels cannot always be directly inferred from the base level, this is because
the detection of features at a certain level is applied among the elements of the space at the previous one, which
already represent abstract features of original data.
Besides being the intermediary representation of an irreducible process of data abstraction, structures at
intermediary levels can be used to trace back the original details of the data and to provide a particular
meaningful interpretation of the space itself.
The peculiar meaning of the space of data at each level is given by the set of features detected at that level, which
should indicate a relational information pattern arising from the removal of redundancies and noise.

\begin{figure}[t]
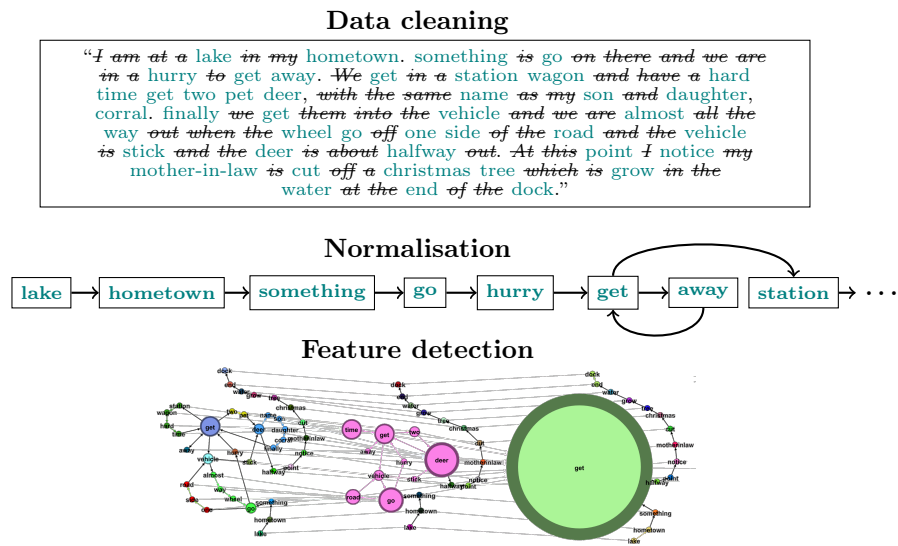

      \centering
      \vspace{-10pt}
      \include{tikzpics/dream_analysis_example}
      \vspace{-20pt}
      \caption{ \small Phases of multilevel graph data analysis on a dream report.}
      \label{fig:dream_analysis}
      \vspace{-10pt}
\end{figure}

Finally, MGDA can draw conclusions about the data by applying the typical quantitative
measures of graph analytics to the various layers of the multilevel graph, which we further discuss in Section~\ref{sec:preliminary-results}.
In addition to the relational aspects, this analysis could also rely on information coming from labels and attributes
of the nodes and edges of each layer, which can be calculated from the base level labels.

\subsection{Feature Detection}

The choice of the features to be detected at each level of the multilevel graph is crucial step in the MGDA process,
which should be based on the domain of the analysis and the specific characteristics of the spaces of data generated by
the normalisation process.
This formally corresponds to the definition of the contraction functions and their order in the $\Gamma$ sequence of
the multilevel graph $M = (G, \Gamma)$.

Examples of features that can be detected in the context of the multilevel graph data analysis are cliques,
cycles, strongly connected components, star formations and other topological structures of interest.
Each of those features can be associated with a particular purpose in elaborating and extracting spatial information.
For instance, cliques, which are complete sub-graphs, don't provide any information about the relationships between the
elements residing in the same clique.
Representing a clique as a supernode in the upper level of the multilevel graph allows the elimination of redundant
information about the relationships between the elements in the clique, highlighting the relationships between the
cliques themselves.
This is especially true if the process of feature detection is combined with a labelling process
that assigns a label to each supernode representing the detected feature, indicating, for example, the number of nodes
in the clique.
Another example is the detection of star formations, which are complete bipartite sub-graphs defined by a central node
connected to other peripheral nodes only adjacent to it.
Merging the peripheral nodes into a supernode allows to highlight the central node as a key element in the space of data,
as making the rest of the nodes dependent on it, hence removing noisy information.

\begin{wrapfigure}{R}{0.5\textwidth}
      \centering
      \vspace{-10pt}
      \begin{tikzpicture}
    [mynode/.style={draw, thick, circle, minimum size=3mm, inner sep=0mm},
    ->,shorten >=1pt,auto,node distance=2cm, thick,main node/.style={circle,draw, minimum size=10mm, inner sep=0mm},
    little node/.style={circle,draw, minimum size=7.5mm, inner sep=0mm},
    every edge/.style={draw, thick, ->}]
    \node[mynode] (1) at (0, -0.25) {};
    \node[mynode] (2) at (0.75, -0.5) {};
    \node[mynode] (3) at (1, 0.25) {};
    \node[mynode] (4) at (0.25, 0.5) {};

    \draw[] (1) -- (2);
    \draw[] (2) -- (3);
    \draw[] (3) -- (4);
    \draw[] (4) -- (1);

    \draw[red, thick, dashed] (2,0) -- (2.5,0);

    \node[mynode, teal, minimum size=5mm] (1c) at (3, 0) {};
\end{tikzpicture}
      \vspace{-30pt}
      \begin{tikzpicture}
    [mynode/.style={draw, thick, circle, minimum size=3mm, inner sep=0mm},
    ->,shorten >=1pt,auto,node distance=2cm, thick,main node/.style={circle,draw, minimum size=10mm, inner sep=0mm},
    little node/.style={circle,draw, minimum size=7.5mm, inner sep=0mm},
    every edge/.style={draw, thick, ->}]
    \node[mynode] (1) at (0, -0.25) {};
    \node[mynode] (2) at (0.75, -0.5) {};
    \node[mynode] (3) at (1, 0.25) {};
    \node[mynode] (4) at (0.25, 0.5) {};

    \node[mynode] (5) at (2.25, 0) {};
    \node[mynode] (6) at (1.5, 0.25) {};
    \node[mynode] (7) at (2, 0.75) {};

    \draw[] (1) -- (2);
    \draw[] (2) -- (3);
    \draw[] (3) -- (4);
    \draw[] (4) -- (1);
    \draw[] (3) -- (6);
    \draw[] (5) -- (6);
    \draw[] (6) -- (7);
    \draw[] (7) -- (5);

    \draw[red, thick,  dashed] (3,0) -- (3.5,0);

    \node[mynode, teal, minimum size=5mm] (1c) at (4, 0) {};
    \node[mynode, teal, minimum size=5mm] (2c) at (5, 0) {};
    \draw[teal] (1c) -- (2c);
\end{tikzpicture}
      \vspace{-30pt}
      \begin{tikzpicture}
    [mynode/.style={draw, thick, circle, minimum size=3mm, inner sep=0mm},
    ->,shorten >=1pt,auto,node distance=2cm, thick,main node/.style={circle,draw, minimum size=10mm, inner sep=0mm},
    little node/.style={circle,draw, minimum size=7.5mm, inner sep=0mm},
    every edge/.style={draw, thick, ->, shorten >=0pt}]
    \node[mynode] (1) at (0, -0.25) {};
    \node[mynode] (2) at (0.75, -0.5) {};
    \node[mynode] (3) at (1, 0.25) {};
    \node[mynode] (4) at (0.25, 0.5) {};
    \node[mynode] (5) at (2, 0) {};
    \node[mynode] (6) at (1.75, 0.75) {};

    \draw[] (1) -- (2);
    \draw[] (2) -- (3);
    \draw[] (3) -- (4);
    \draw[] (4) -- (1);
    \draw[] (5) -- (6);
    \draw[] (3) -- (5);
    \draw[] (6) -- (3);

    \draw[red, thick, dashed] (2.5,0) -- (3,0);

    \node[mynode, teal, minimum size=5mm] (1c) at (3.5, 0) {};
    \node[mynode, teal, minimum size=5mm] (2c) at (4.5, 0) {};
    \node[mynode, teal, minimum size=5mm] (3c) at (5.5, 0) {};
    \draw[every edge, thick, teal, line width=1.5pt] (1c) -- (2c);
    \draw[every edge, thick, teal, line width=1.5pt] (2c) -- (1c);
    \draw[every edge, thick, teal, line width=1.5pt] (2c) -- (3c);
    \draw[every edge, thick, teal, line width=1.5pt] (3c) -- (2c);
\end{tikzpicture}
      \vspace{-20pt}
      \caption{\small Examples of spaces of data arising from simple cycle detection on graphs.}
      \label{fig:space_example}
      \vspace{-10pt}
\end{wrapfigure}
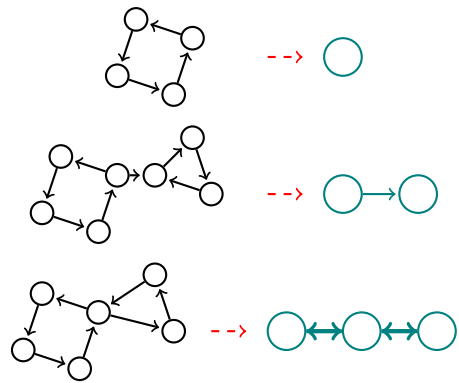

The detection of these features can be performed by applying specific algorithms to the space of data, which can
identify the presence of those patterns of interest and provide a representation of them in terms of supernodes.
The complexity of such algorithms, which can reach the exponential order, is still an open problem that can be addressed
with the aid of high-performance computing solutions or by applying heuristics to reduce the computational cost.

This is especially true, for instance, for patterns such as cliques and simple cycles, which also share the peculiar
property of not necessarily defining a partition of the nodes upon detection. 
Since the detection of these features aims to describe the topological structure of the space of data, best results
can be achieved by applying generic contraction algorithms that produce supernodes based on the common
memberships of the nodes in the detected features.
In other words, if we see the product of a contraction function $f_{C_i}$ as quotient graph $G_i$ of the original one
$G_{i-1}$ in input, where supernodes in $G_i$ correspond to the equivalence classes of the nodes in $G_{i-1}$,
then the equivalence relation which originates that quotient graph $G_i$ could be defined as follows:
let $u, v$ be nodes in $G_{i-1}$, let $Q = \{F_1, F_2, \ldots, F_n\}$ be the set of features detected in $G_{i-1}$,
let $\sim$ be the equivalence relation which aggregates the nodes in the same supernode in $G_{i}$, then:
\begin{equation*}
    u \sim v \quad \Leftrightarrow \quad \exists S \subseteq Q \quad \text{ s.t. } \quad \forall F \in S, \; u, v \in F \quad \land \quad \forall F \in \{Q \setminus S\}, \; u, v \notin F
\end{equation*}

\section{Preliminary Results of MGDA Approach} \label{sec:preliminary-results}

In this section, we discuss preliminary results about the benefits of applying the typical measurements of
graph analytics in combination with multilevel graph data analysis, by providing examples of quantitative
parameters that can be computed on the various layers of the multilevel graph and by considering their implications and
meaning in the context of the data space analysis.
For each considered parameter, we provide a brief description of the related results in the context of
dram narrative analysis, which is the scenario we used to test the proposed approach as described in
Section~\ref{sec:dream-narrative-analysis}.

\subsection{Assortativity Coefficient}

Assortativity is a measure of the tendency of nodes to connect to other nodes that are similar according to
a numerical attribute.
Considering the nature of the multilevel graph, assortativity can be computed on attributes that are assigned to
nodes of the graph at a certain level.
Those attributes carry information that can only be derived from both the labels assigned to the nodes in the base
level and the subsequently detected features in the process of abstraction.
This means that measures like the assortativity coefficient of a graph at a certain level can provide insightful
information about the quality of relationships between upper-level points, which can be used to infer the presence
of hidden patterns in the original space of data.

For instance, in our study applied to dream narratives analysis, we computed the assortativity coefficient on the
weight attribute of the nodes in the various layers of the multilevel graph.
This attribute was assigned to each node based on the number of occurrences of the original words which contributed to
the formation of that node, recursively calculated upon weights of the nodes in the previous layer.
Results showed that assortativity coefficient of graphs in the multilevel graph varied considerably, both among
levels of abstraction, indicating that each layer revealed different hidden patterns in the space of data, both
among dreamers, characterising the different nature of the space in terms of trend through levels.

\begin{figure}
      \centering
      \vspace{-20pt}
      \includegraphics[width=0.8\textwidth]{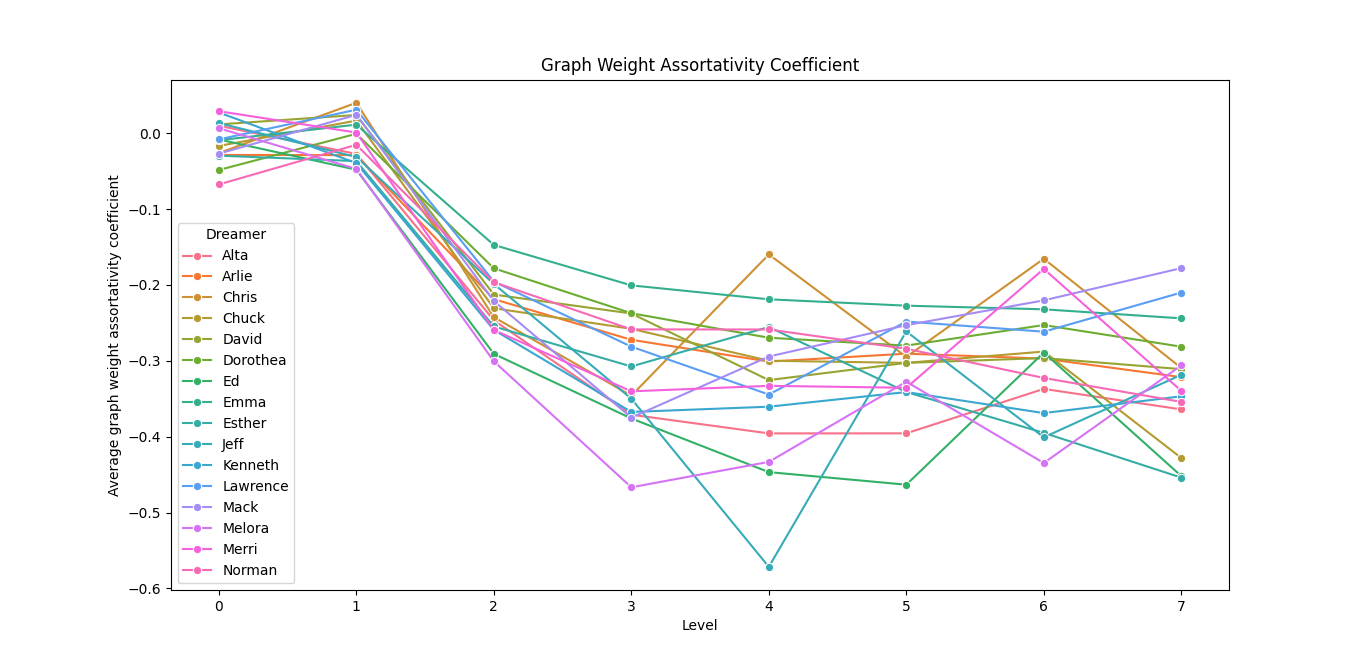}
      \vspace{-10pt}
      \caption{\small Average assortativity coefficients among levels for the $P$ group of dreamers.}
        \label{fig:assortativity_example}
        \vspace{-25pt}
\end{figure}

\begin{wrapfigure}{r}{0.5\textwidth}
      \centering
      \vspace{-31pt}
      \includegraphics[width=0.5\textwidth]{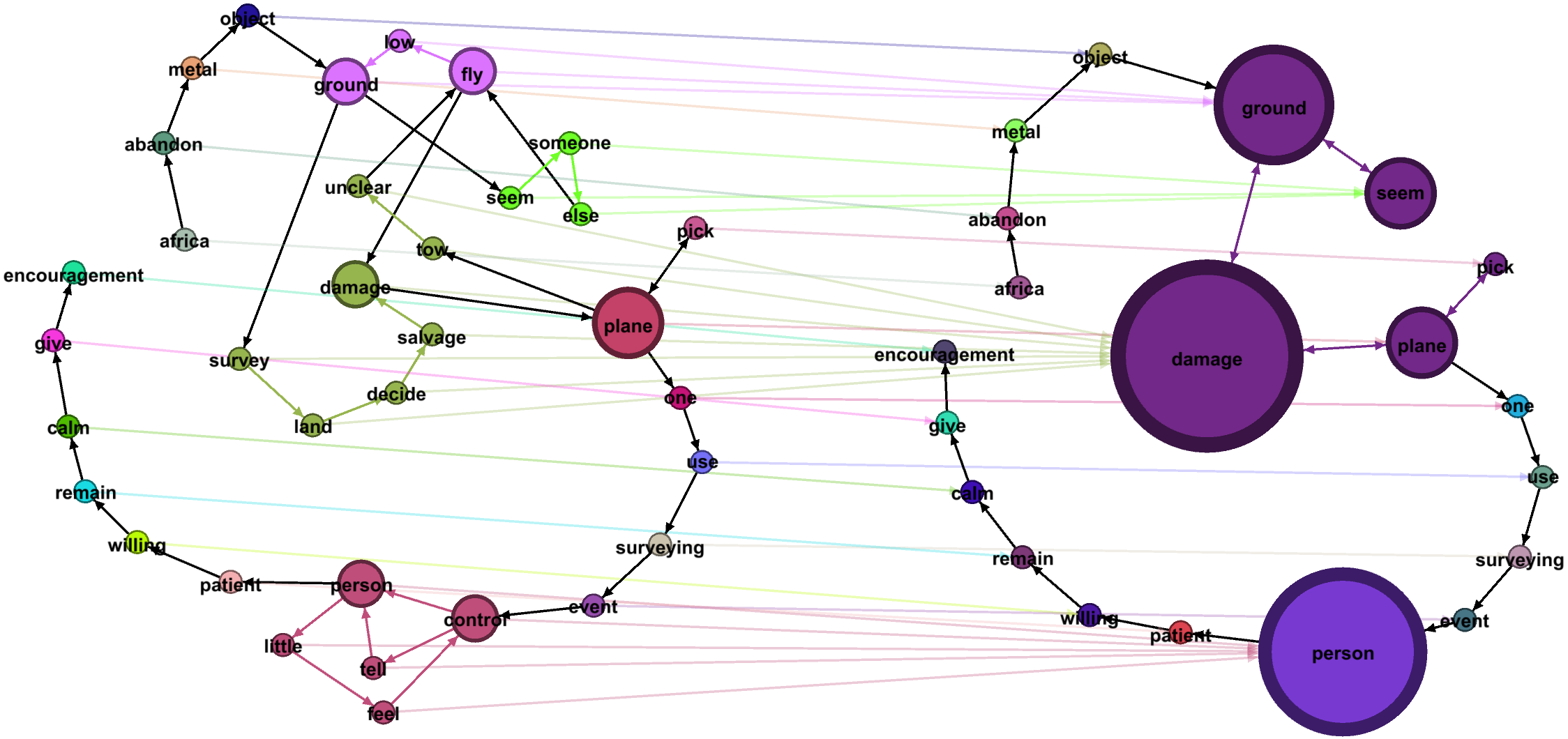}
      \caption{\small David's dream contraction by simple cycles (58.8\% contraction percentage).}
      \label{fig:contracton_graphs_example}
      \vspace{-30pt}
\end{wrapfigure}

\subsection{Contraction Percentage}

The contraction percentage is a measure of the amount of redundant or noisy information that is removed from the
space of data during the process of abstraction, which can be computed at each level $i > 0$ by comparing the number of
nodes at that $i^\text{th}$ level with the number of supernodes in the $(i-1)^\text{th}$ level.

This measure can provide a quantitative insight into the quality of the abstraction process while remaining
strictly related to the specific features detected at each level and the levels below.
This means that the specific meaning of the percentage of contraction can be inferred from the chosen contraction
functions and the order in which they are applied.

\begin{figure}
      \centering
      \vspace{-20pt}
      \includegraphics[width=0.8\textwidth]{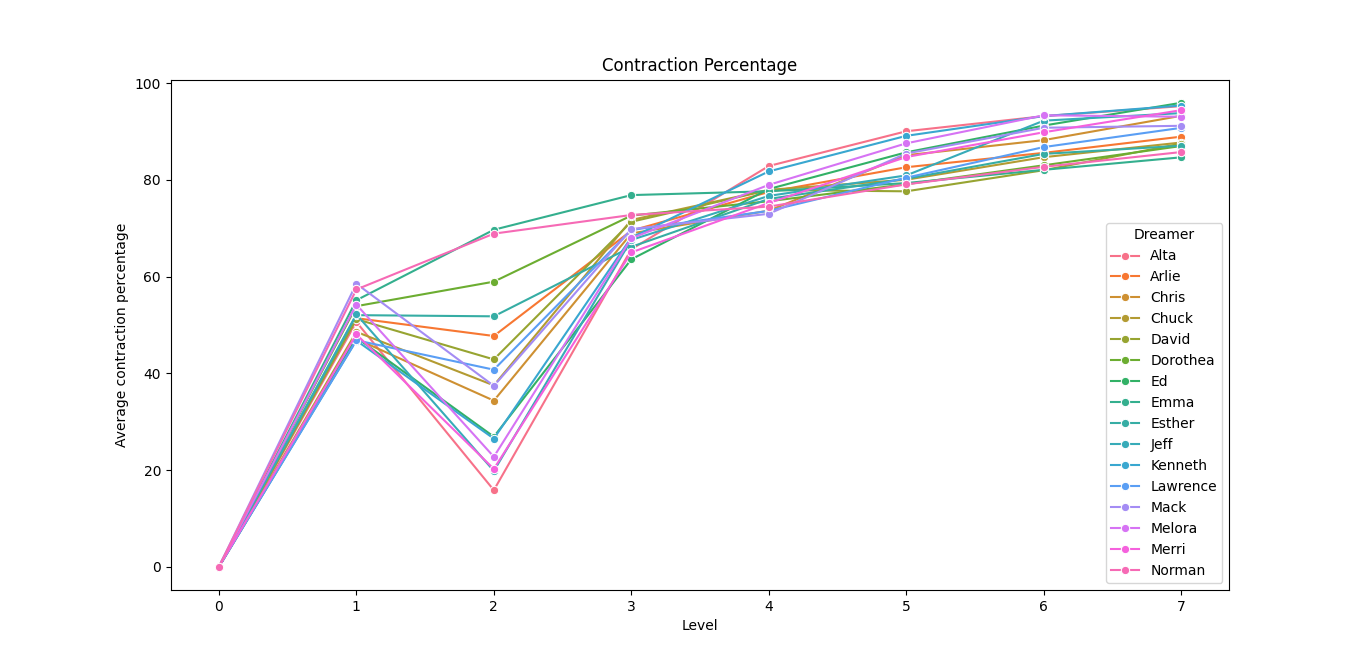}
      \vspace{-10pt}
      \caption{\small Average contraction percentage among levels for the $P$ group of dreamers.}
        \label{fig:contraction_example}
        \vspace{-15pt}
\end{figure}

The results in Figure~\ref{fig:contraction_example} reveals that spaces of words derived from dream narratives of
individuals can be effectively abstracted and distinguished by applying strongly connected component contractions.
This allows us to highlight the general connectivity among words, which can be interpreted as a general coherence among
topics in the same narratives.
Results in our study also show that, besides being a crucial aspect of the data space of dreams, this
could help in making macro-distinctions among dreamers, that exhibited different average levels of coherence in their narratives.
In general, this parameter could help to characterise the fundamentally different nature of spaces and provide keys to
understanding its consequences on the upper levels.

\subsection{Graph Density}

Graph density is a measure of the number of edges in a graph compared to the total number of possible edges according
to the number of nodes.
In the context of the multilevel graph, computing graph density at each level of the hierarchy and observing its trend,
can provide insightful information about the level of strength of relationships between the elements in the space,
according to the detected features and the impact of the single steps of the abstraction process.
In general, having low density in the upper levels of the multilevel graph can be interpreted as a sign of a simpler,
concise and more linear representation of the space of data.
Those aspects could also be evaluated to infer what contraction criteria effectively reduce dimensionality and
complexity from a topological point of view.
In other cases, an increase in density can be interpreted as a sign of summarization of the core relationships between
points of the space, which could omit linear and redundant relationships and highlight the fundamental complex patterns
among groups of points.

\begin{figure}
      \centering
      \vspace{-20pt}
      \includegraphics[width=0.8\textwidth]{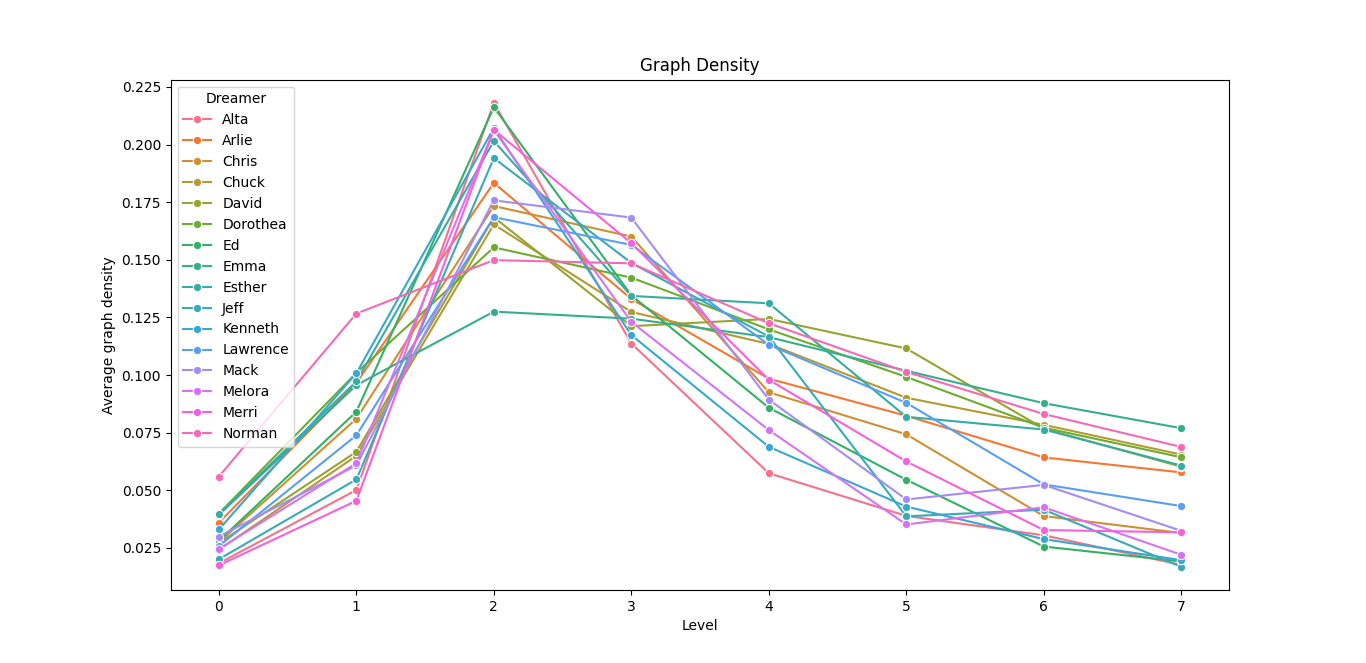}
      \vspace{-10pt}
      \caption{\small Avarage graph density among levels for the $P$ group of dreamers.}
        \label{fig:density_example}
        \vspace{-15pt}
\end{figure}

In our study on dreams, we computed the average density at each level of the multilevel graph for each dreamer, as
shown in Figure~\ref{fig:density_example}.
What we observed, in general, is that the density of the graphs tended to increase in the first levels of the
hierarchy, followed by a decrease in the subsequent levels.
This suggests that the first steps of the abstraction process, such as cycle and strongly connected component detection,
tended to reduce linear redundant information, highlighting the core relationships between the elements in the data
space, which could also be attributed to the contexts of words in the same narratives.
The trend is also informative about the linearity of the structures starting from level 2.
Indeed, steps of abstraction subsequently applied after that level, in this case star formations detection, tended to
make the representation of the space of data more concise, by removing noisy information and highlighting the main areas of
interest in the space.
Another interesting aspect is also the diversity of trends among different dreamers.
This suggests that the effects of contraction on density are tied to the nature of the space itself,
which could have distinctive characteristics among individuals.

\section{Dream Narratives Analysis} \label{sec:dream-narrative-analysis}

This section provides more detailed insights into the study conducted on dream narrative collections using the MGDA approach.
The data in this domain typically comprises written or transcribed reports of dreams, often supplemented by metadata
such as the dreamer's demographics and the date of the dream.
The primary goal of applying MGDA is to identify recurring patterns of thought that emerge through word usage and to
uncover typical correlations between thoughts and contexts within individual dreams.

In this study, a multilevel graph was constructed for each dream: {\it supernodes} represent individual words at the
lowest level, while at higher levels they assume the role of groups of words, concepts, and thoughts.
The {\it superedges} capture the relationships between those groups of words, inferred from the nearness property of original words.
The contracted multilevel graph provides a global view, representing the emerging behaviours of an individual dreamer
or a group of dreamers.

\subsection{Dream Dataset}

During the study, we analysed a dataset of dream narratives from a group $P$ of sixteen individuals,
and a group $T$ of adolescents. All data is taken from the DreamBank repository ({\it https://dreambank.net/}).

We applied the described MGDA methodology, with a preprocessing phase for each dream report consisting of text cleaning,
lemmatization, stopwords removal, unigram creation, and the construction of a directed graph representing unigram sequentiality relations.
The resulting dream-related graph $G$ is used as input to the {\it natural transformation function $\eta$}, in order to
construct the graph at level 0, formally $G_0$.

Constructing the multilevel graph structure allowed us to gather data for observing
the evolution of the graph topology through parameters changing at different levels.\\

The study can be divided into two phases related to portions of the dataset:
\begin{enumerate}[1.]
\item In the first phase we analysed the group $P$.
    This group was characterized by a heterogeneous set of dreamers in terms of sex and age, as well as a diverse
    length of periods in which the dreams were recorded.
    Three particular dreamers among the original sixteen -- Arlie (A), Emma (E), and Norman (N) -- were taken as
    representatives of the different behaviours observed in the sample of sixteen dreamers.
    Those three dreamers were selected based on the high number of dreams available in the dataset.
    This part of the analysis aims to observe whether it is possible that average behaviours can change
    considerably among different individuals when taken singularly. \\
\item In the second phase we analyzed the group $T$.
    We compared group $P$ with group $T$, both considered as single entities, to observe the behaviour of a group with
    a wide age range concerning a group of individuals with a smaller, more specific age range ($14$ to $18$ years old).
    In this case, single individuals in the group $P$ were equally weighted in the number of considered narratives, so
    $30$ random dreams for each dreamer were selected, for a total amount of $480$ dreams for group $P$.
    This part of the analysis aims to observe whether aspects having psychological impact, such as age, have a
    common influence on the average shape of the dream-related space of data.
\end{enumerate}

\subsection{Detected Features}\label{subsec:detected-features}

Following the creation of the $G$ graph and the construction of $G_0$, the analysis of cycles and other formations
(shapes) is described by the following detections:
\begin{itemize}
\item {\bf simple cycles},  performed by applying the  contracting function $f_{C_1}$;
\item {\bf strongly connected components}   performed by applying the contracting function $f_{C_2}$;
\item {\bf star formations},  performed by applying the subsequent contraction functions $f_{C_3} \dots f_{C_7}$.
\end{itemize}

\begin{figure}
    \centering
    \vspace{-15pt}
    \includegraphics[width=\textwidth]{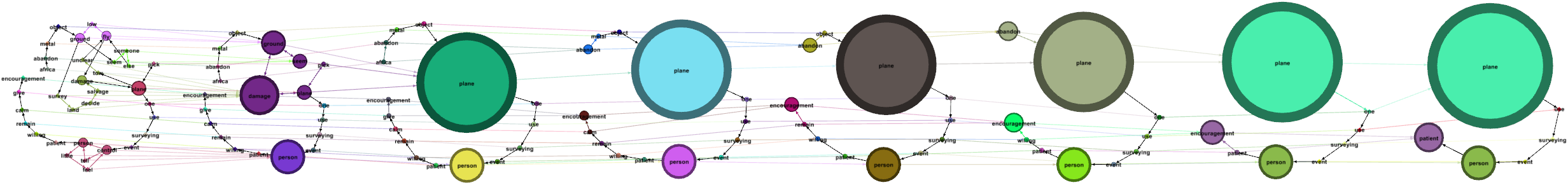}
    \caption{\small Contraction process of a multilevel graph originated by a dream narrative.}
    \label{fig:dream_all_levels}
    \vspace{-15pt}
\end{figure}

Simple cycles could represent circular patterns of thoughts in the dream narratives, as being closed loops of never
repeating words, while strongly connected components -- subgraphs of mutually reachable nodes -- indicate clusters of
closely interrelated and thematic coherent concepts or thoughts that co-occur and interconnect within the dreams.
Star formations represent a subgraph where a specific node has the centralised control, indicating that a word or a
thought persists in any relation.

\subsection{Results}

In this section, we present the results of the analysis conducted on the dream narratives of group $P$ and group $T$.
In Table~\ref{tab:average_measures}, we provide a summary of the average measures gathered in the two phases of the study.
In particular, $P$ and $T$ represent equally weighted groups of dreamers, while $P_c$ represents the
typical behaviour of the examined group $P = \{P_1. P_2, \ldots, P_{16}\}$.
Each of its considered average property $k_i(P_c)$, with $k \in \{NNC, CP, \ldots, GDe\}$ representing the measurement,
and $i \in \{0, \ldots, 7\}$ representing the level where the measurement was taken, can be formally defined as $k_i(P_c) = \frac{1}{|P|} \sum_{j=1}^{|P|} k_i(P_j)$.

Each of the presented charts comprises trajectories representing the average behaviour in a pool
of dreams related to a single dreamer or a group of individuals.
For each pool, we selected dreams between 15 and 300 lemma occurrences per narrative, roughly corresponding to a
dream length ranging from 30 to 750 words, to avoid the influence of outliers while maintaining a
large sample size of dreams of typical length.
The charts show the evolution of a certain property across progressive levels associated with different contraction
schemes as described in Subsection~\ref{subsec:detected-features}.
The ordinate axis quantifies the average property value, while the abscissa delineates multilevel graph levels.
Due to the significant variation in the average lemma count among dreamers' narratives — for instance, 38.57 for Emma,
28.80 for Norman, and 48.47 for Arlie — the data sensitive to dream length has been normalised to a percentage value
relative to the lemma count of the original dream.

\paragraph{Phase 1}

\begin{figure}[h]
    \begin{subfigure}{0.32\textwidth}
        \includegraphics[width=1.1\textwidth]{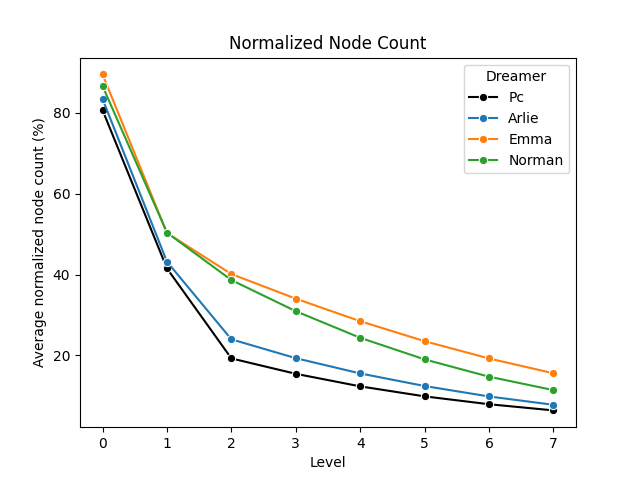}
        \caption{\scriptsize NNC}\label{fig:node_count}
    \end{subfigure} \hfill
    \begin{subfigure}{0.32\textwidth}
        \includegraphics[width=1.1\textwidth]{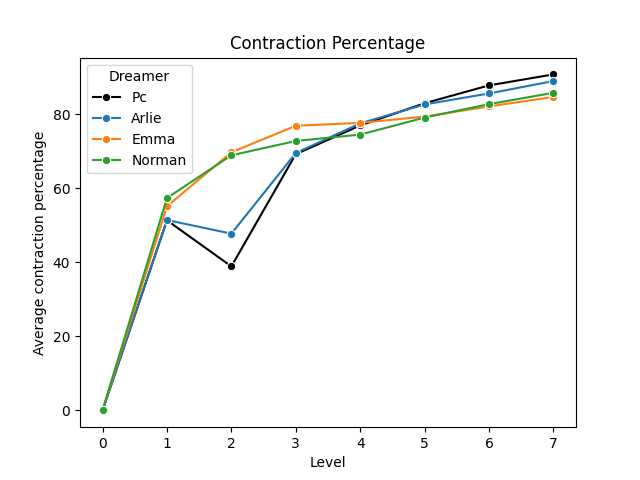}
        \caption{\scriptsize CP}\label{fig:contraction_percentage}
    \end{subfigure} \hfill
    \begin{subfigure}{0.32\textwidth}
        \includegraphics[width=1.1\textwidth]{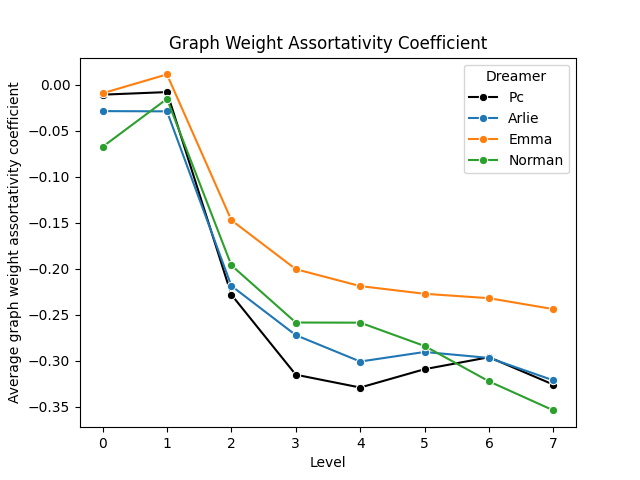}
        \caption{\scriptsize GWAC}\label{fig:assortativity_coefficient}
    \end{subfigure} \hfill
    \begin{subfigure}{0.32\textwidth}
        \includegraphics[width=1.1\textwidth]{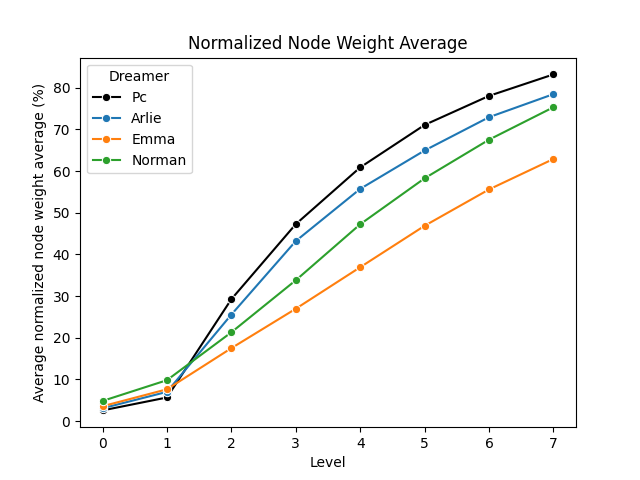}
        \caption{\scriptsize NNW}\label{fig:node_weight_average}
    \end{subfigure} \hfill
    \begin{subfigure}{0.32\textwidth}
        \includegraphics[width=1.1\textwidth]{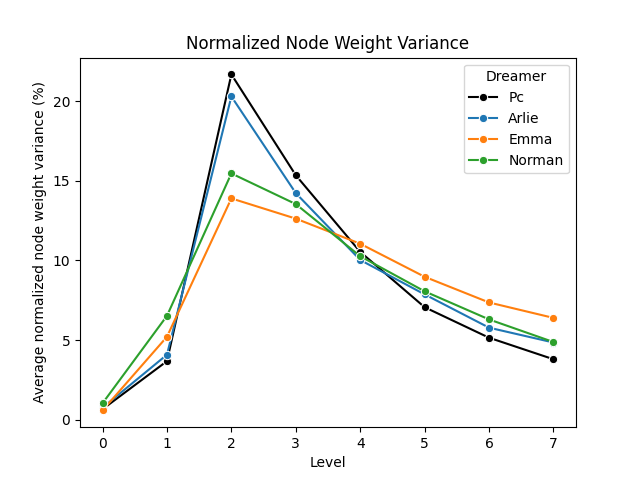}
        \caption{\scriptsize NNWv}\label{fig:node_weight_variance}
    \end{subfigure} \hfill
    \begin{subfigure}{0.32\textwidth}
        \includegraphics[width=1.1\textwidth]{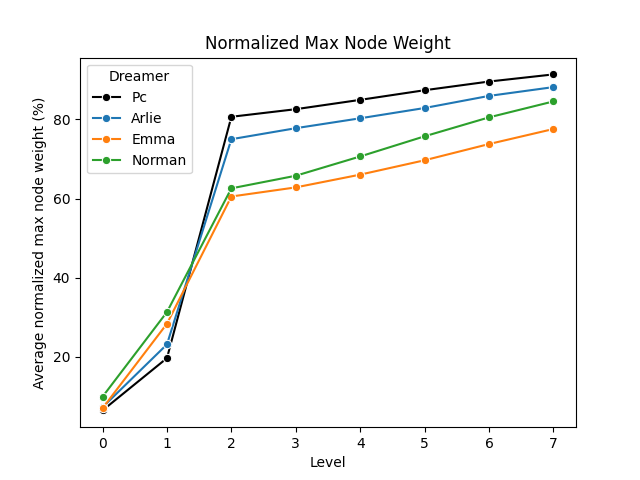}
        \caption{\scriptsize NNWm}\label{fig:max_node_weight}
    \end{subfigure}
    \begin{subfigure}{0.32\textwidth}
        \includegraphics[width=1.1\textwidth]{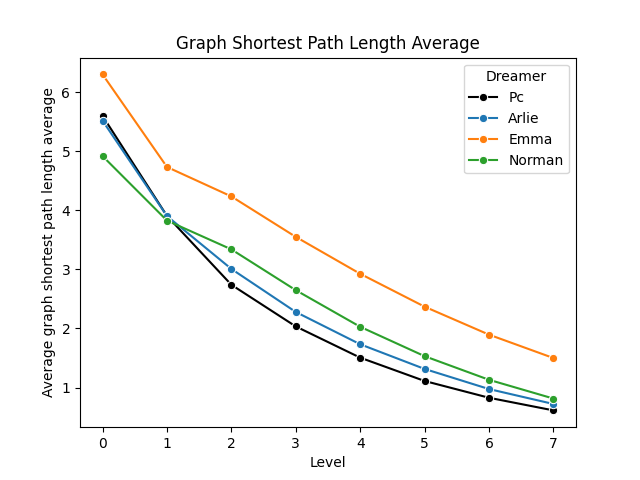}
        \caption{\scriptsize GSPL}\label{fig:shortest_path_length_average}
    \end{subfigure} \hfill
    \begin{subfigure}{0.32\textwidth}
        \includegraphics[width=1.1\textwidth]{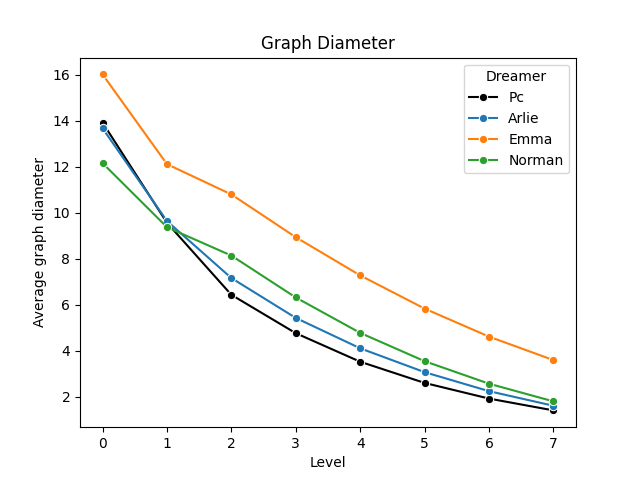}
        \caption{\scriptsize GDi}\label{fig:graph_diameter}
    \end{subfigure} \hfill
    \begin{subfigure}{0.32\textwidth}
        \includegraphics[width=1.1\textwidth]{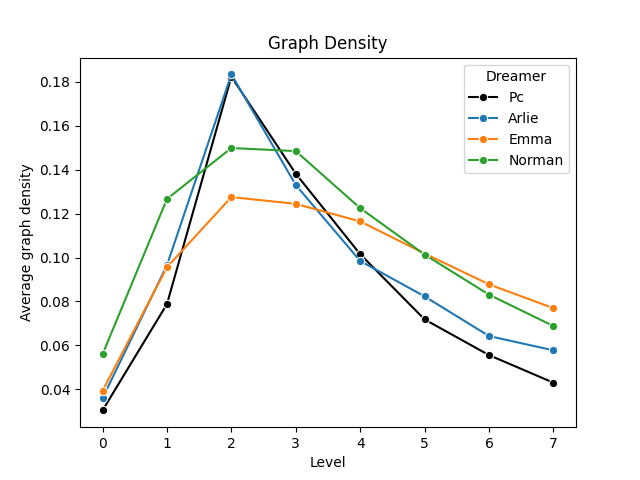}
        \caption{\scriptsize GDe}\label{fig:graph_density}
    \end{subfigure}
    \caption{\small Statistics on phase 1 of the study.}\label{fig:emma_norman_arlie}
    \vspace{-15pt}
\end{figure}

Charts presented in Fig.~\ref{fig:emma_norman_arlie} comprise three distinct trajectories, each corresponding to the
oneiric data of the three selected dreamers plus the average behaviour of $P_c$.

\newpage

\begin{table}[H]
\centering
\renewcommand{\arraystretch}{0.94}
\begin{tabular}{|c|c|c|c|c|c|c|c|c|c|c|c|c|}
\cline{2-13}
\multicolumn{1}{c|}{} & {\scriptsize \bf Records} & {\scriptsize \bf Age} & {\scriptsize \bf Lv.} & {\scriptsize \bf NNC} & {\scriptsize \bf CP} & {\scriptsize \bf GWAC}  & {\scriptsize \bf NNW} & {\scriptsize \bf NNWv} & {\scriptsize \bf NNWm} & {\scriptsize \bf GSPL} & {\scriptsize \bf GDi} & {\scriptsize \bf GDe}\\
\hline
\multicolumn{13}{|c|}{1$^\text{st}$ Phase} \\
\hline
\multirow{8}*{\scriptsize \bf $P_c$} & \multirow{8}*{\footnotesize 5869} & \multirow{8}*{\makecell{\tiny Widely \\ \tiny ranged}} &
       \scriptsize \textbf0 &\scriptsize 80.63 &\scriptsize 0.000 &\scriptsize -0.0107 &\scriptsize 2.558 &\scriptsize 0.6565 &\scriptsize 6.454 &\scriptsize 5.596 &\scriptsize 13.92 &\scriptsize 0.0304 \\ \cline{4-13}
 & & & \scriptsize \textbf1 &\scriptsize 41.51 &\scriptsize 51.35 &\scriptsize -0.0079 &\scriptsize 5.657 &\scriptsize 3.675 &\scriptsize 19.67 &\scriptsize 3.904 &\scriptsize 9.566 &\scriptsize 0.0788 \\ \cline{4-13}
 & & & \scriptsize \textbf2 &\scriptsize 19.26 &\scriptsize 38.89 &\scriptsize -0.2283 &\scriptsize 29.25 &\scriptsize 21.69 &\scriptsize 80.69 &\scriptsize 2.739 &\scriptsize 6.436 &\scriptsize 0.1823 \\ \cline{4-13}
 & & & \scriptsize \textbf3 &\scriptsize 15.46 &\scriptsize 69.21 &\scriptsize -0.3153 &\scriptsize 47.25 &\scriptsize 15.35 &\scriptsize 82.61 &\scriptsize 2.034 &\scriptsize 4.770 &\scriptsize 0.1381 \\ \cline{4-13}
 & & & \scriptsize \textbf4 &\scriptsize 12.36 &\scriptsize 76.95 &\scriptsize -0.3292 &\scriptsize 60.87 &\scriptsize 10.54 &\scriptsize 84.97 &\scriptsize 1.505 &\scriptsize 3.531 &\scriptsize 0.1015 \\ \cline{4-13}
 & & & \scriptsize \textbf5 &\scriptsize 9.882 &\scriptsize 82.96 &\scriptsize -0.3093 &\scriptsize 71.06 &\scriptsize 7.064 &\scriptsize 87.40 &\scriptsize 1.111 &\scriptsize 2.610 &\scriptsize 0.0718 \\ \cline{4-13}
 & & & \scriptsize \textbf6 &\scriptsize 7.938 &\scriptsize 87.78 &\scriptsize -0.2962 &\scriptsize 78.03 &\scriptsize 5.147 &\scriptsize 89.60 &\scriptsize 0.8271 &\scriptsize 1.931 &\scriptsize 0.0556 \\ \cline{4-13}
 & & & \scriptsize \textbf7 &\scriptsize 6.413 &\scriptsize 90.75 &\scriptsize -0.3261 &\scriptsize 83.19 &\scriptsize 3.795 &\scriptsize 91.40 &\scriptsize 0.6133 &\scriptsize 1.418 &\scriptsize 0.0430 \\
\hline
\multirow{8}*{\scriptsize \bf A} & \multirow{8}*{\footnotesize 221} & \multirow{8}*{\makecell{\tiny Middle \\ \tiny aged}} &
       \scriptsize \textbf0 & \scriptsize 83.39 & \scriptsize 0.000 & \scriptsize -0.0286 & \scriptsize 3.066 & \scriptsize 0.7486 & \scriptsize 7.244 & \scriptsize 5.515 & \scriptsize 13.67 & \scriptsize 0.0360 \\ \cline{4-13}
 & & & \scriptsize \textbf1 & \scriptsize 43.17 & \scriptsize 51.41 & \scriptsize -0.0289 & \scriptsize 7.008 & \scriptsize 4.082 & \scriptsize 23.14 & \scriptsize 3.901 & \scriptsize 9.639 & \scriptsize 0.0965 \\ \cline{4-13}
 & & & \scriptsize \textbf2 & \scriptsize 23.98 & \scriptsize 47.74 & \scriptsize -0.2187 & \scriptsize 25.55 & \scriptsize 20.30 & \scriptsize 75.02 & \scriptsize 3.007 & \scriptsize 7.165 & \scriptsize 0.1834 \\ \cline{4-13}
 & & & \scriptsize \textbf3 & \scriptsize 19.32 & \scriptsize 69.51 & \scriptsize -0.2722 & \scriptsize 43.20 & \scriptsize 14.23 & \scriptsize 77.80 & \scriptsize 2.278 & \scriptsize 5.433 & \scriptsize 0.1330 \\ \cline{4-13}
 & & & \scriptsize \textbf4 & \scriptsize 15.55 & \scriptsize 77.52 & \scriptsize -0.3010 & \scriptsize 55.72 & \scriptsize 10.03 & \scriptsize 80.31 & \scriptsize 1.732 & \scriptsize 4.113 & \scriptsize 0.0984 \\ \cline{4-13}
 & & & \scriptsize \textbf5 & \scriptsize 12.44 & \scriptsize 82.63 & \scriptsize -0.2906 & \scriptsize 64.93 & \scriptsize 7.870 & \scriptsize 82.91 & \scriptsize 1.314 & \scriptsize 3.077 & \scriptsize 0.0823 \\ \cline{4-13}
 & & & \scriptsize \textbf6 & \scriptsize 9.858 & \scriptsize 85.60 & \scriptsize -0.2970 & \scriptsize 72.92 & \scriptsize 5.778 & \scriptsize 85.97 & \scriptsize 0.9742 & \scriptsize 2.253 & \scriptsize 0.0643 \\ \cline{4-13}
 & & & \scriptsize \textbf7 & \scriptsize 7.797 & \scriptsize 88.95 & \scriptsize -0.3215 & \scriptsize 78.46 & \scriptsize 4.846 & \scriptsize 88.19 & \scriptsize 0.7216 & \scriptsize 1.619 & \scriptsize 0.0578 \\
\cline{2-13}
\multirow{8}*{\scriptsize \bf E} & \multirow{8}*{\footnotesize 1252} & \multirow{8}*{\makecell{\tiny Widely \\ \tiny ranged}} &
       \scriptsize \textbf0 & \scriptsize 89.47 & \scriptsize 0.000 & \scriptsize -0.0092 & \scriptsize 3.564 & \scriptsize 0.5812 & \scriptsize 7.009 & \scriptsize 6.298 & \scriptsize 16.01 & \scriptsize 0.0393 \\ \cline{4-13}
 & & & \scriptsize \textbf1 & \scriptsize 50.25 & \scriptsize 55.13 & \scriptsize  0.0113 & \scriptsize 7.622 & \scriptsize 5.208 & \scriptsize 28.37 & \scriptsize 4.732 & \scriptsize 12.11 & \scriptsize 0.0956 \\ \cline{4-13}
 & & & \scriptsize \textbf2 & \scriptsize 40.12 & \scriptsize 69.70 & \scriptsize -0.1472 & \scriptsize 17.50 & \scriptsize 13.90 & \scriptsize 60.53 & \scriptsize 4.234 & \scriptsize 10.80 & \scriptsize 0.1275 \\ \cline{4-13}
 & & & \scriptsize \textbf3 & \scriptsize 34.03 & \scriptsize 76.87 & \scriptsize -0.2005 & \scriptsize 26.95 & \scriptsize 12.63 & \scriptsize 62.84 & \scriptsize 3.549 & \scriptsize 8.938 & \scriptsize 0.1244 \\ \cline{4-13}
 & & & \scriptsize \textbf4 & \scriptsize 28.46 & \scriptsize 77.67 & \scriptsize -0.2189 & \scriptsize 36.92 & \scriptsize 11.05 & \scriptsize 66.06 & \scriptsize 2.925 & \scriptsize 7.279 & \scriptsize 0.1164 \\ \cline{4-13}
 & & & \scriptsize \textbf5 & \scriptsize 23.52 & \scriptsize 79.36 & \scriptsize -0.2273 & \scriptsize 46.86 & \scriptsize 8.976 & \scriptsize 69.69 & \scriptsize 2.368 & \scriptsize 5.838 & \scriptsize 0.1018 \\ \cline{4-13}
 & & & \scriptsize \textbf6 & \scriptsize 19.25 & \scriptsize 82.10 & \scriptsize -0.2321 & \scriptsize 55.58 & \scriptsize 7.361 & \scriptsize 73.80 & \scriptsize 1.895 & \scriptsize 4.620 & \scriptsize 0.0878 \\ \cline{4-13}
 & & & \scriptsize \textbf7 & \scriptsize 15.59 & \scriptsize 84.68 & \scriptsize -0.2440 & \scriptsize 62.89 & \scriptsize 6.392 & \scriptsize 77.60 & \scriptsize 1.499 & \scriptsize 3.601 & \scriptsize 0.0769 \\
\cline{2-13}
\multirow{8}*{\scriptsize \bf N} & \multirow{8}*{\footnotesize 669} & \multirow{8}*{\tiny Unknown} &
       \scriptsize \textbf0 & \scriptsize 86.58 & \scriptsize 0.000 & \scriptsize -0.0675 & \scriptsize 4.826 & \scriptsize 1.029 & \scriptsize 9.915 & \scriptsize 4.917 & \scriptsize 12.16 & \scriptsize 0.0560 \\ \cline{4-13}
 & & & \scriptsize \textbf1 & \scriptsize 50.35 & \scriptsize 57.35 & \scriptsize -0.0153 & \scriptsize 9.810 & \scriptsize 6.522 & \scriptsize 31.32 & \scriptsize 3.826 & \scriptsize 9.374 & \scriptsize 0.1267 \\ \cline{4-13}
 & & & \scriptsize \textbf2 & \scriptsize 38.63 & \scriptsize 68.90 & \scriptsize -0.1963 & \scriptsize 21.30 & \scriptsize 15.47 & \scriptsize 62.60 & \scriptsize 3.338 & \scriptsize 8.140 & \scriptsize 0.1498 \\ \cline{4-13}
 & & & \scriptsize \textbf3 & \scriptsize 32.57 & \scriptsize 76.58 & \scriptsize -0.2364 & \scriptsize 30.83 & \scriptsize 13.82 & \scriptsize 65.12 & \scriptsize 2.764 & \scriptsize 6.747 & \scriptsize 0.1384 \\ \cline{4-13}
 & & & \scriptsize \textbf4 & \scriptsize 27.17 & \scriptsize 77.95 & \scriptsize -0.2536 & \scriptsize 40.27 & \scriptsize 11.86 & \scriptsize 68.38 & \scriptsize 2.276 & \scriptsize 5.558 & \scriptsize 0.1310 \\ \cline{4-13}
 & & & \scriptsize \textbf5 & \scriptsize 22.39 & \scriptsize 80.13 & \scriptsize -0.2610 & \scriptsize 49.46 & \scriptsize 9.815 & \scriptsize 72.06 & \scriptsize 1.850 & \scriptsize 4.541 & \scriptsize 0.1093 \\ \cline{4-13}
 & & & \scriptsize \textbf6 & \scriptsize 18.32 & \scriptsize 82.35 & \scriptsize -0.2623 & \scriptsize 57.55 & \scriptsize 8.224 & \scriptsize 75.53 & \scriptsize 1.500 & \scriptsize 3.681 & \scriptsize 0.0909 \\ \cline{4-13}
 & & & \scriptsize \textbf7 & \scriptsize 14.82 & \scriptsize 84.83 & \scriptsize -0.2778 & \scriptsize 63.78 & \scriptsize 7.014 & \scriptsize 78.53 & \scriptsize 1.214 & \scriptsize 2.981 & \scriptsize 0.0746 \\
\hline
\multicolumn{13}{|c|}{2$^\text{nd}$ Phase} \\
\hline
\multirow{8}*{\scriptsize \bf P} & \multirow{8}*{\footnotesize 480}  & \multirow{8}*{\makecell{\tiny Widely \\ \tiny ranged}} &
       \scriptsize \textbf0 & \scriptsize 80.90 & \scriptsize 0.000 & \scriptsize -0.0033 & \scriptsize 2.564 & \scriptsize 0.638 & \scriptsize 6.437 & \scriptsize 5.652 & \scriptsize 14.11 & \scriptsize 0.0303 \\ \cline{4-13}
 & & & \scriptsize \textbf1 & \scriptsize 42.01 & \scriptsize 51.78 & \scriptsize -0.0024 & \scriptsize 5.956 & \scriptsize 3.758 & \scriptsize 20.01 & \scriptsize 3.997 & \scriptsize 9.867 & \scriptsize 0.0764 \\ \cline{4-13}
 & & & \scriptsize \textbf2 & \scriptsize 20.36 & \scriptsize 41.01 & \scriptsize -0.1909 & \scriptsize 26.63 & \scriptsize 22.24 & \scriptsize 79.50 & \scriptsize 2.897 & \scriptsize 6.848 & \scriptsize 0.1871 \\ \cline{4-13}
 & & & \scriptsize \textbf3 & \scriptsize 16.44 & \scriptsize 68.20 & \scriptsize -0.3034 & \scriptsize 45.42 & \scriptsize 15.58 & \scriptsize 81.53 & \scriptsize 2.165 & \scriptsize 5.119 & \scriptsize 0.1417 \\ \cline{4-13}
 & & & \scriptsize \textbf4 & \scriptsize 13.27 & \scriptsize 76.46 & \scriptsize -0.2832 & \scriptsize 59.75 & \scriptsize 10.09 & \scriptsize 84.02 & \scriptsize 1.612 & \scriptsize 3.840 & \scriptsize 0.0954 \\ \cline{4-13}
 & & & \scriptsize \textbf5 & \scriptsize 10.76 & \scriptsize 83.72 & \scriptsize -0.2830 & \scriptsize 68.79 & \scriptsize 7.564 & \scriptsize 86.17 & \scriptsize 1.226 & \scriptsize 2.900 & \scriptsize 0.0775 \\ \cline{4-13}
 & & & \scriptsize \textbf6 & \scriptsize 8.694 & \scriptsize 87.11 & \scriptsize -0.2437 & \scriptsize 76.42 & \scriptsize 5.048 & \scriptsize 88.68 & \scriptsize 0.917 & \scriptsize 2.173 & \scriptsize 0.0538 \\ \cline{4-13}
 & & & \scriptsize \textbf7 & \scriptsize 7.090 & \scriptsize 91.11 & \scriptsize -0.2859 & \scriptsize 80.98 & \scriptsize 3.881 & \scriptsize 90.36 & \scriptsize 0.698 & \scriptsize 1.631 & \scriptsize 0.0411 \\
\hline
\multirow{8}*{\scriptsize \bf T} & \multirow{8}*{\footnotesize 169} & \multirow{8}*{\makecell{\tiny $14 \leq$ \\ \tiny Age \\ \tiny $\leq 18$}} &
       \scriptsize \textbf0 & \scriptsize 82.80 & \scriptsize 0.000 & \scriptsize -0.0335 & \scriptsize 3.895 & \scriptsize 1.003 & \scriptsize 8.796 & \scriptsize 4.894 & \scriptsize 11.80 & \scriptsize 0.0465 \\ \cline{4-13}
 & & & \scriptsize \textbf1 & \scriptsize 42.61 & \scriptsize 51.27 & \scriptsize -0.0290 & \scriptsize 8.591 & \scriptsize 6.238 & \scriptsize 29.10 & \scriptsize 3.499 & \scriptsize 8.450 & \scriptsize 0.1185 \\ \cline{4-13}
 & & & \scriptsize \textbf2 & \scriptsize 26.03 & \scriptsize 54.25 & \scriptsize -0.2347 & \scriptsize 25.41 & \scriptsize 20.32 & \scriptsize 75.59 & \scriptsize 2.781 & \scriptsize 6.497 & \scriptsize 0.1772 \\ \cline{4-13}
 & & & \scriptsize \textbf3 & \scriptsize 20.16 & \scriptsize 69.01 & \scriptsize -0.3565 & \scriptsize 41.79 & \scriptsize 16.22 & \scriptsize 77.91 & \scriptsize 2.073 & \scriptsize 4.740 & \scriptsize 0.1576 \\ \cline{4-13}
 & & & \scriptsize \textbf4 & \scriptsize 15.45 & \scriptsize 73.22 & \scriptsize -0.2894 & \scriptsize 58.36 & \scriptsize 9.949 & \scriptsize 81.21 & \scriptsize 1.465 & \scriptsize 3.355 & \scriptsize 0.1043 \\ \cline{4-13}
 & & & \scriptsize \textbf5 & \scriptsize 11.94 & \scriptsize 82.11 & \scriptsize -0.3860 & \scriptsize 67.54 & \scriptsize 8.309 & \scriptsize 84.54 & \scriptsize 1.071 & \scriptsize 2.373 & \scriptsize 0.0955 \\ \cline{4-13}
 & & & \scriptsize \textbf6 & \scriptsize 9.080 & \scriptsize 83.86 & \scriptsize -0.2657 & \scriptsize 77.09 & \scriptsize 6.186 & \scriptsize 88.06 & \scriptsize 0.740 & \scriptsize 1.592 & \scriptsize 0.0845 \\ \cline{4-13}
 & & & \scriptsize \textbf7 & \scriptsize 7.086 & \scriptsize 88.13 & \scriptsize -0.4920 & \scriptsize 84.63 & \scriptsize 3.831 & \scriptsize 91.58 & \scriptsize 0.489 & \scriptsize 1.065 & \scriptsize 0.0475 \\
\hline
\end{tabular}
\vspace{10pt}
\caption{\small Table of the average measures gathered in the two phases of the study.}
\label{tab:average_measures}
\end{table}

Figure~\ref{fig:node_count} shows node count evolution, highlighting faster simplification in Arlie's data and more gradual
reduction in Norman and Emma's.
Figure~\ref{fig:contraction_percentage} compares node contraction percentages across levels, revealing significant
structural divergence for Arlie.
Figures~{\ref{fig:max_node_weight},~\ref{fig:node_weight_average} and~\ref{fig:node_weight_variance}} display the
distribution of node weights, where Arlie's data exhibits larger node weights, indicating higher aggregation, while Emma and Norman show more balanced growth.
Figure~\ref{fig:shortest_path_length_average} tracks the evolution of the average shortest path, where Arlie’s data
simplifies fastest, while Norman and Emma follow similar trends.
Figure~\ref{fig:graph_diameter} shows graph diameter reduction, which correlates with path contraction.
Figure~\ref{fig:graph_density} depicts increasing graph density through contraction, with Arlie’s showing the fastest
densification and Norman's and Emma's converging in upper levels.
Figure~\ref{fig:assortativity_coefficient} measures the tendency of nodes with similar weights to connect,
with Norman showing declining assortativity, while Emma maintains more stable, interconnected aggregations.

\begin{figure}[h]
    \begin{subfigure}{0.32\textwidth}
        \includegraphics[width=1.1\textwidth]{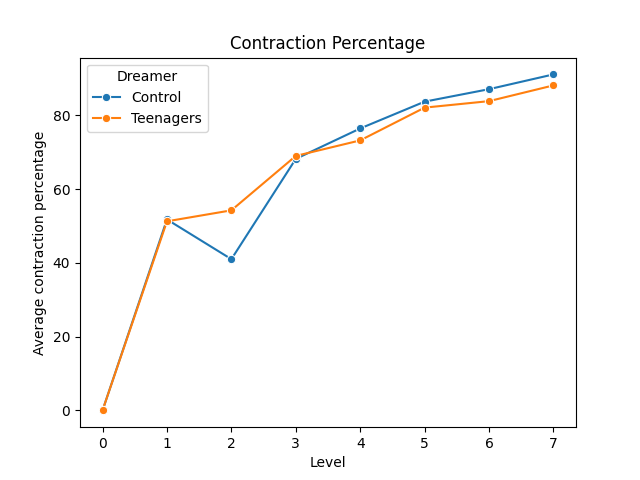}
        \caption{\scriptsize CP}\label{fig:contraction_percentage_teens}
    \end{subfigure} \hfill
    \begin{subfigure}{0.32\textwidth}
        \includegraphics[width=1.1\textwidth]{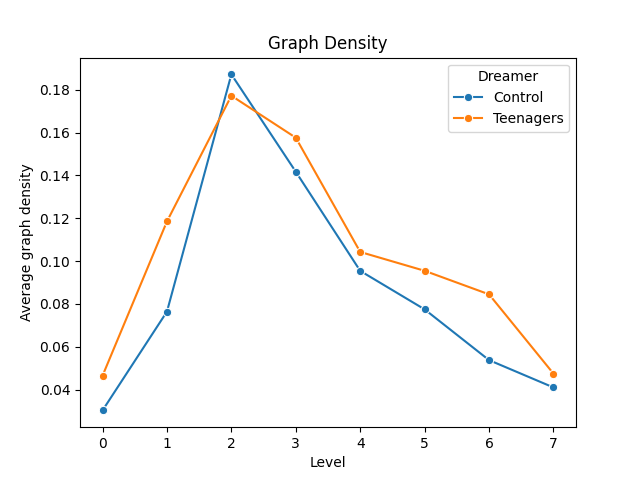}
        \caption{\scriptsize GDe}\label{fig:density_teens}
    \end{subfigure} \hfill
    \begin{subfigure}{0.32\textwidth}
        \includegraphics[width=1.1\textwidth]{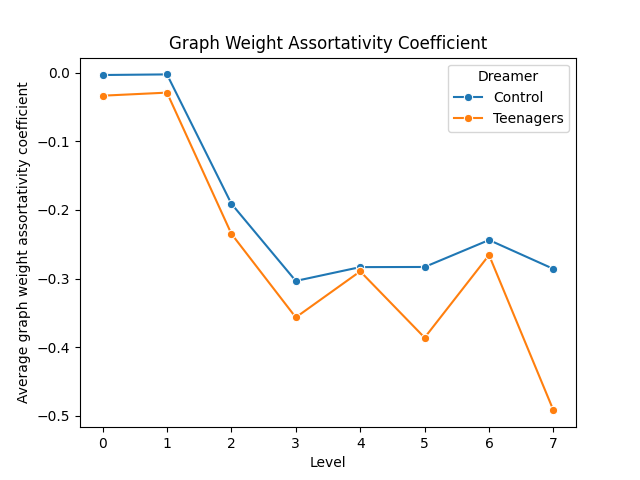}
        \caption{\scriptsize GWAC}\label{fig:assortativity_coefficient_teens}
    \end{subfigure} \hfill
    \begin{subfigure}{0.32\textwidth}
        \includegraphics[width=1.1\textwidth]{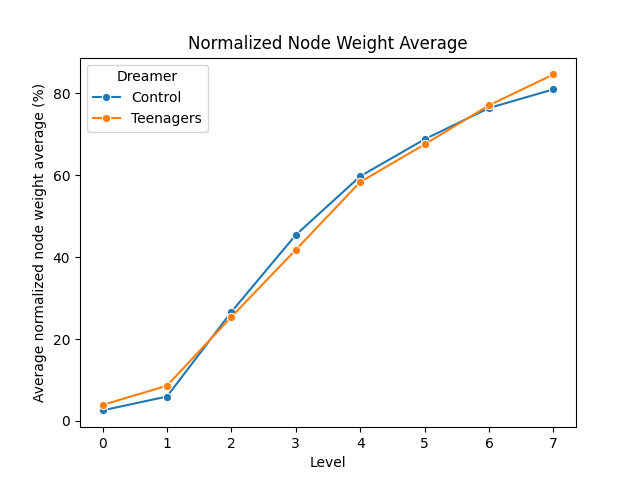}
        \caption{\scriptsize NNW}\label{fig:normalized_node_weight_teens}
    \end{subfigure} \hfill
    \begin{subfigure}{0.32\textwidth}
        \includegraphics[width=1.1\textwidth]{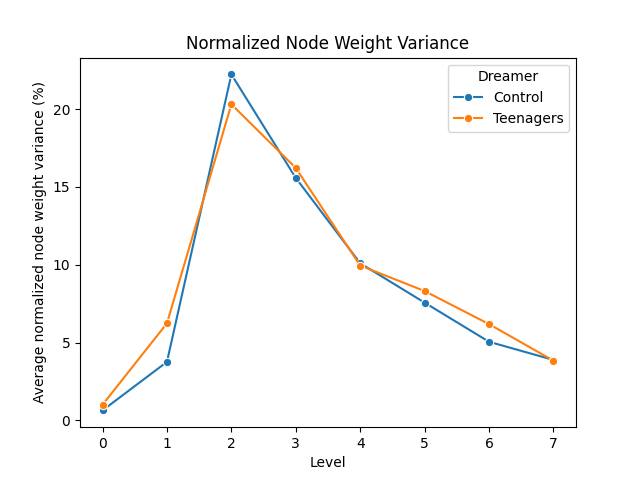}
        \caption{\scriptsize NNWv}\label{fig:normalized_node_weight_variance_teens}
    \end{subfigure}
    \begin{subfigure}{0.32\textwidth}
        \includegraphics[width=1.1\textwidth]{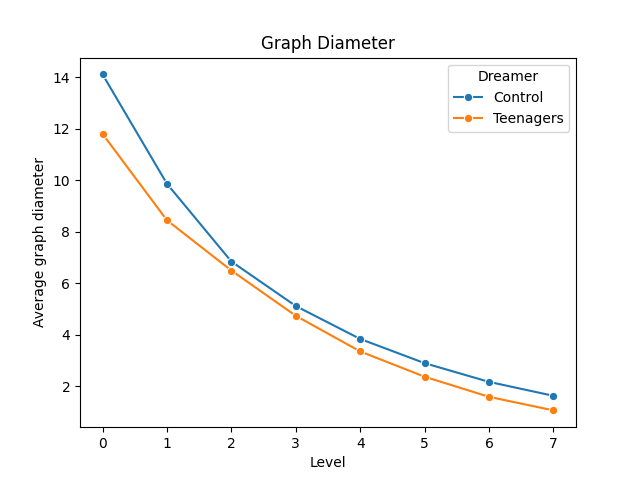}
        \caption{\scriptsize GDi}\label{fig:graph_diameter_teens}
    \end{subfigure} \hfill
    \caption{\small Statistics on phase 2 of the study.}\label{fig:teens}
    \vspace{-15pt}
\end{figure}

In Fig.~\ref{fig:teens} is presented the analysis of the group of adolescents.
In this case, trajectories represent the average behaviour of a group of individuals, where each of those
is equally weighted in the number of considered narratives.

The contraction percentage chart in Figure~\ref{fig:contraction_percentage_teens} shows that while both groups behave
similarly, at the beginning and end, adolescents experience more fragmentation during strongly connected component
contraction.
In Figure~\ref{fig:density_teens}, the graph density chart reveals that adolescents have a higher overall density, suggesting more
interconnectedness. Anyway, nodes remain unaggregated and poorly connected during the strongly connected component process.
The weight assortativity coefficient chart in Figure~\ref{fig:assortativity_coefficient_teens} indicates greater
instability in adolescents, especially in later levels, suggesting fluctuations in node importance and tree-like
connections between contexts.

Besides the considered statistics, the overall results show that the group of adolescents shares similar characteristics
with the control group, as evidenced by the strong similarity between the behaviour illustrated in
Figures~{\ref{fig:normalized_node_weight_teens}, \ref{fig:normalized_node_weight_variance_teens} and~\ref{fig:graph_diameter_teens}}.
Other aspects not shown in this section, such as the statistics regarding normalized node and edge counts, edge weights,
shortest path lengths, contraction set sizes and cycle basis counts, also exhibit very similar trends between the two
groups.

\paragraph{Phase 2}

\section{Conclusions and Future Work}
In this work, we introduced the concept of the multilevel graph as a data structure representing a \textit{dataspace} at
different levels of abstraction with concern for traceability.
This could serve as a tool for modelling and managing data spaces, providing a common representation of both
data sources and the data itself, and act as a uniform navigation medium, providing a unified way to
model unstructured formats of data. Future works in this direction could focus on the possibility of representing
structured sources of data as decontractible and, to some extent, ``queryable'' graphs.

We have also introduced the Multilevel Graph Data Analysis (MGDA) approach as a novel method for analyzing complex
\textit{dataspaces}, particularly those derived from unstructured data sources. By leveraging a multilevel graph structure,
MGDA allows for the systematic abstraction of data through the sequential application of data cleaning, normalization,
and feature detection.
The key advantage of MGDA lies in its ability to represent raw data as increasingly abstract relational structures
while preserving critical topological information at each level of the analysis according to the particular features
of interest, facilitating the discovery of hidden patterns and providing multiple views of the space of data arising from
the original information.

The results of applying MGDA in specific scenarios, such as the analysis of dream narratives, demonstrate its potential
to reveal underlying relational trends across different levels of abstraction.
Metrics like assortativity, contraction percentage, and graph density provide quantitative insights into the strength
and complexity of these relationships, enabling a more robust understanding of the data's topological structure.

Furthermore, the preliminary results of the analysis of dream narratives show that MGDA can effectively capture the
distinctive characteristics of spaces of data representing individuals or groups of individuals, providing a selective
tool for identifying dreamers that present atypical behaviour based on quantitative parameters.
This aspect could be better explored in future studies, by combining dream narratives with more detailed metadata and
considering possible correlations between those aspects and the demographic, psychological or psychiatric traits of the
dreamers.

\section{\ackname}
This work has been partially funded by the European Union - NextGenerationEU under the Italian Ministry of University
and Research (MUR) National Innovation Ecosystem grant ECS00000041 - VITALITY.
We thank Lorenzo Lupi for his contribution in defining and developing the algorithms supporting the management of the
multilevel graph data structure.

%
%

\bibliographystyle{splncs04}

%

\end{document}